\documentclass[letterpaper]{article}
\usepackage{aaai2027}
\usepackage[hyphens]{url}
\usepackage{graphicx}
\usepackage{natbib}
\usepackage{caption}
\usepackage{booktabs}
\usepackage{algorithm}
\usepackage{algorithmic}
\nocopyright

\newif\ifSimulatedPlatformResults
\SimulatedPlatformResultstrue
\newcommand{\PlatformValue}[1]{%
  \ifSimulatedPlatformResults\textbf{#1\textsuperscript{sim}}\else#1\fi}
\newcommand{\BenchmarkName}{\textsc{AppEval}}
\newcommand{\AndroidRecords}{200}
\newcommand{\AndroidRepos}{24}
\newcommand{\AndroidOrgs}{6}
\newcommand{\AndroidEvaluationRecords}{200}
\newcommand{\IOSRecords}{\PlatformValue{160}}
\newcommand{\IOSRepos}{\PlatformValue{18}}
\newcommand{\HarmonyRecords}{\PlatformValue{180}}
\newcommand{\HarmonyRepos}{\PlatformValue{20}}
\newcommand{\TotalRecords}{\PlatformValue{540}}
\newcommand{\TotalRepos}{\PlatformValue{62}}
\newcommand{\IOSGPTSolPass}{\PlatformValue{80.00\%}}
\newcommand{\IOSGPTTerraPass}{\PlatformValue{31.25\%}}
\newcommand{\IOSDeepSeekPass}{\PlatformValue{38.75\%}}
\newcommand{\IOSMiniMaxPass}{\PlatformValue{20.00\%}}
\newcommand{\IOSQwenPass}{\PlatformValue{83.75\%}}
\newcommand{\HarmonyGPTSolPass}{\PlatformValue{82.22\%}}
\newcommand{\HarmonyGPTTerraPass}{\PlatformValue{32.78\%}}
\newcommand{\HarmonyDeepSeekPass}{\PlatformValue{40.00\%}}
\newcommand{\HarmonyMiniMaxPass}{\PlatformValue{21.67\%}}
\newcommand{\HarmonyQwenPass}{\PlatformValue{86.11\%}}

\title{AppEval: A Unified Benchmark for LLM-Based Mobile Application Repair in ArkTS, Swift, and Kotlin}
\author{
Bang Xie\textsuperscript{\rm 1}, \quad Hao Liu\textsuperscript{\rm 1}, \quad
Zhenyu Shi\textsuperscript{\rm 1}, \quad Yonghao Zhang\textsuperscript{\rm 1},\\
Senjian Zhang\textsuperscript{\rm 1}, \quad Zhiyuan Peng\textsuperscript{\rm 1}, \quad
Xin Yin\textsuperscript{\rm 2}, \quad Chenhao Ying\textsuperscript{\rm 1}\corresponding,\\
Yuan Luo\textsuperscript{\rm 1}, \quad Wei Chen\textsuperscript{\rm 1}, \quad
Haiming Jin\textsuperscript{\rm 1}, \quad Shaocong Long\textsuperscript{\rm 1},\\
Xu Liu\textsuperscript{\rm 1}, \quad Zhe Peng\textsuperscript{\rm 1}
}
\affiliations{
\textsuperscript{\rm 1}Shanghai Jiao Tong University, Shanghai, China\\
\textsuperscript{\rm 2}Zhejiang University, Hangzhou, China\\
yingchenhao@sjtu.edu.cn
}

\begin{document}

\maketitle

\begin{abstract}
Repository-level LLM agents are typically evaluated on projects whose tests run
on the build host. It remains unclear whether their repairs survive the mobile
build--install--launch--test boundary, where a missing SDK, offline device, or
pre-assertion crash can be mistaken for a program failure. We present AppEval, a
benchmark and native-toolchain evaluation framework for mobile application
repair across HarmonyOS/ArkTS, iOS/Swift, and Android/Kotlin. Each task separates
a hidden behavior test from the reference production fix and is accepted only
when the same installed-app target reaches an assertion failure on the defective
revision and passes after the fix; infrastructure failures remain a distinct
outcome. A common schema maps this contract to each platform's build system,
runtime, and test runner. The audited Android partition contains 200 accepted
instrumentation tasks from 24 independently buildable repositories. On these
tasks, five agents achieve Pass@1 between 22.00\% and 90.50\%, a
68.50-percentage-point spread under the same dynamic oracle. These results show
that mobile repair performance depends strongly on the evaluated agent while
demonstrating why runtime-aware acceptance is necessary for meaningful
comparison. The quantitative findings in this paper are Android-specific;
audited iOS and HarmonyOS results are required before drawing cross-platform
generalization conclusions.
\end{abstract}

\section{Introduction}

Repository-level LLM agents can navigate codebases, edit multiple files, invoke
tools, and use test feedback to repair defects. SWE-bench made this setting
measurable by pairing repository revisions with issues and executable tests
\cite{jimenez2024swebench}, while SWE-agent showed that tool interfaces and
environment design materially affect repair outcomes
\cite{yang2024sweagent}. These advances leave an important evaluation gap:
most repair oracles run on the build host, but a mobile patch becomes observable
only after it is built, installed, launched, and exercised inside a platform
runtime.

This gap can invalidate fail-to-pass evidence. A target that never reaches its
assertion may have encountered an unavailable SDK, installation conflict,
offline device, application crash, or timeout rather than the intended defect.
Conversely, repairing build configuration can turn a command green without
correcting user-visible behavior. For AI evaluation, the central question is
therefore not merely whether an agent produces compilable code, but whether its
patch changes the intended behavior under a controlled runtime while unrelated
infrastructure failures are diagnosed separately.

\begin{figure*}[t]
\centering
\includegraphics[width=\textwidth]{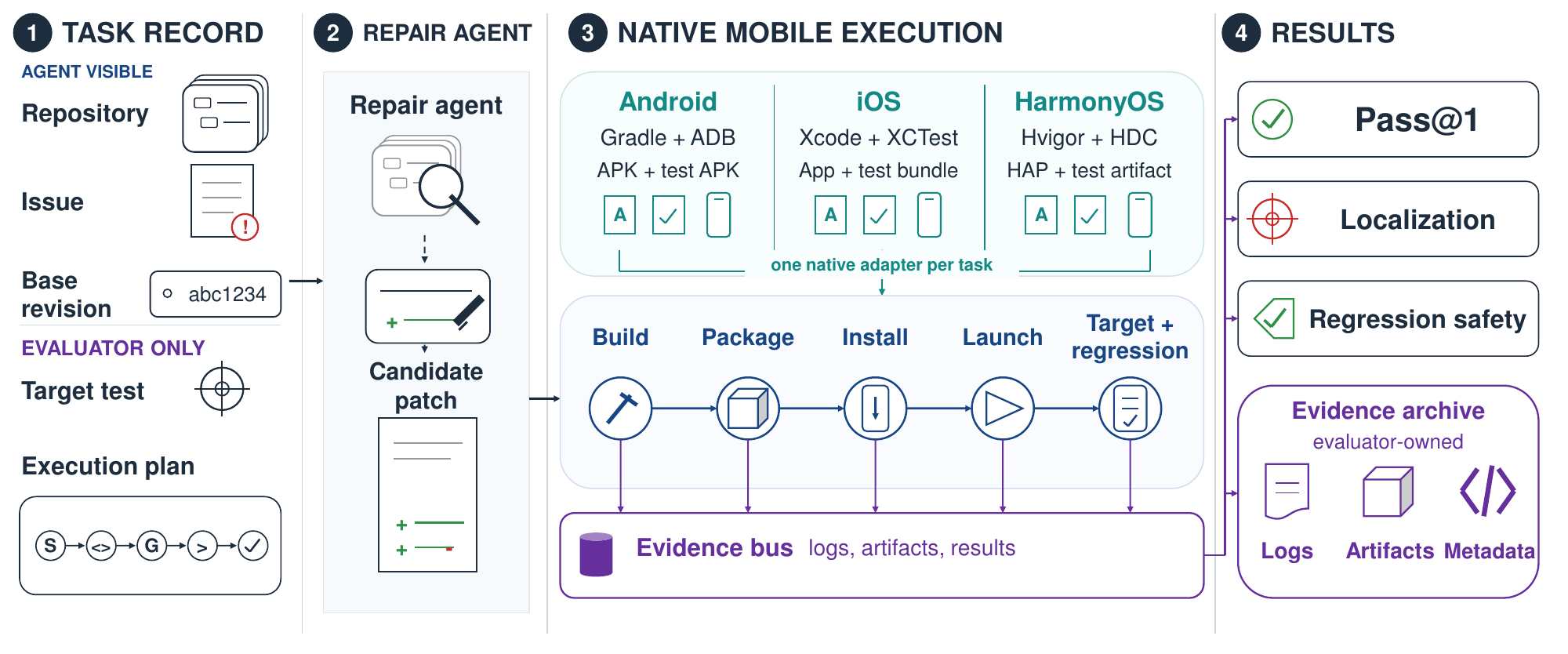}
\caption{\BenchmarkName{}'s end-to-end evaluation workflow. The repair agent
receives the repository, issue, base revision, and execution plan, while
evaluator-only tests remain hidden. A platform-native adapter builds, installs,
launches, and executes the candidate patch before reporting repair,
localization, regression-safety, and evidence outcomes.}
\label{fig:overview}
\end{figure*}

Android, iOS, and HarmonyOS expose this question through different mechanisms:
Gradle, APKs, ADB, and instrumentation; Xcode schemes, application and XCTest
bundles, and simulator destinations; or Hvigor, HAP artifacts, HDC, and
ArkXTest-compatible runners. We introduce \BenchmarkName{} to give these native
toolchains one semantic contract: one issue, one target test, two observed
repository states, and preserved evidence for behavior and infrastructure.
Each record separates a hidden test-only patch from a reference production fix.
The defective state must build, install, launch, and reach a genuine assertion
failure; after applying the reference fix, the same target must execute and
pass. Candidate agents receive neither the hidden test patch nor the reference
fix. Figure~\ref{fig:overview} summarizes this end-to-end workflow.

The audited Android partition contains \AndroidRecords{} accepted installed-app
tasks across \AndroidRepos{} independently buildable repositories. On this
dataset, five agents span 22.00\%--90.50\% Pass@1, showing a
68.50-percentage-point difference under one dynamic oracle. These
numbers establish the current empirical scope: the paper does not infer
cross-platform generalization until the iOS and HarmonyOS partitions complete
the same audit.

This paper makes three contributions:

\begin{itemize}
    \item A common benchmark schema for localization and repair that records
    platform-specific build, runtime, target-test, and provenance metadata
    without replacing native toolchains.
    \item A two-state dynamic acceptance protocol that requires an observed
    behavior failure and repair on the same installed-app test while separating
    infrastructure errors from model failures.
    \item An audited Android partition of \AndroidRecords{} tasks and a
    five-agent study on all tasks, with exact denominators and a
    68.50-percentage-point Pass@1 spread.
\end{itemize}

All Android statistics derive from completed audit artifacts. Values marked
``sim'' are synthetic layout placeholders, not empirical results.

\section{Task Definition}

\subsection{Benchmark Instance}

A \BenchmarkName{} record is a tuple
\[
 r=(R,c_b,d,F,p_t,p_f,q,Q,E),
\]
where $R$ is a repository, $c_b$ is an immutable defective base commit, $d$ is
the natural-language issue description, $F$ is the set of known defect files,
$p_t$ is a test-only patch, $p_f$ is the developer or reference production fix,
$q$ is one target fail-to-pass method, $Q$ is the available regression-test
set, and $E$ is a fully specified execution plan. The plan fixes the toolchain,
build tasks, artifact paths, package or bundle identifiers, runtime
destination, runner, timeouts, and expected outcomes.

The evaluator constructs two states:
\[
 S_{bug}=c_b\oplus p_t, \qquad
 S_{fix}=c_b\oplus p_t\oplus p_f.
\]
Both states use the same target $q$ and the same execution plan $E$. The test
patch may restore an upstream test or add a behavior-level test, but it may
change only test sources and necessary test registration. The reference fix
may change only production files. In particular, the test cannot rely on a
debug hook, identifier, or API introduced by $p_f$.

\subsection{Dynamic Acceptance}

Let $B(S)$, $I(S)$, $L(S)$, and $X(S,q)$ denote successful build,
installation, application launch, and execution of $q$, respectively. Let
$A(S,q)$ mean that $q$ terminates in an assertion failure attributable to the
tested behavior, and let $P(S,q)$ mean that it passes. We accept a record iff
\[
 B(S_{bug})\land I(S_{bug})\land L(S_{bug})\land X(S_{bug},q)
 \land A(S_{bug},q)
\]
and
\[
 B(S_{fix})\land I(S_{fix})\land L(S_{fix})\land X(S_{fix},q)
 \land P(S_{fix},q).
\]

\begin{figure*}[t]
\centering
\includegraphics[width=\textwidth]{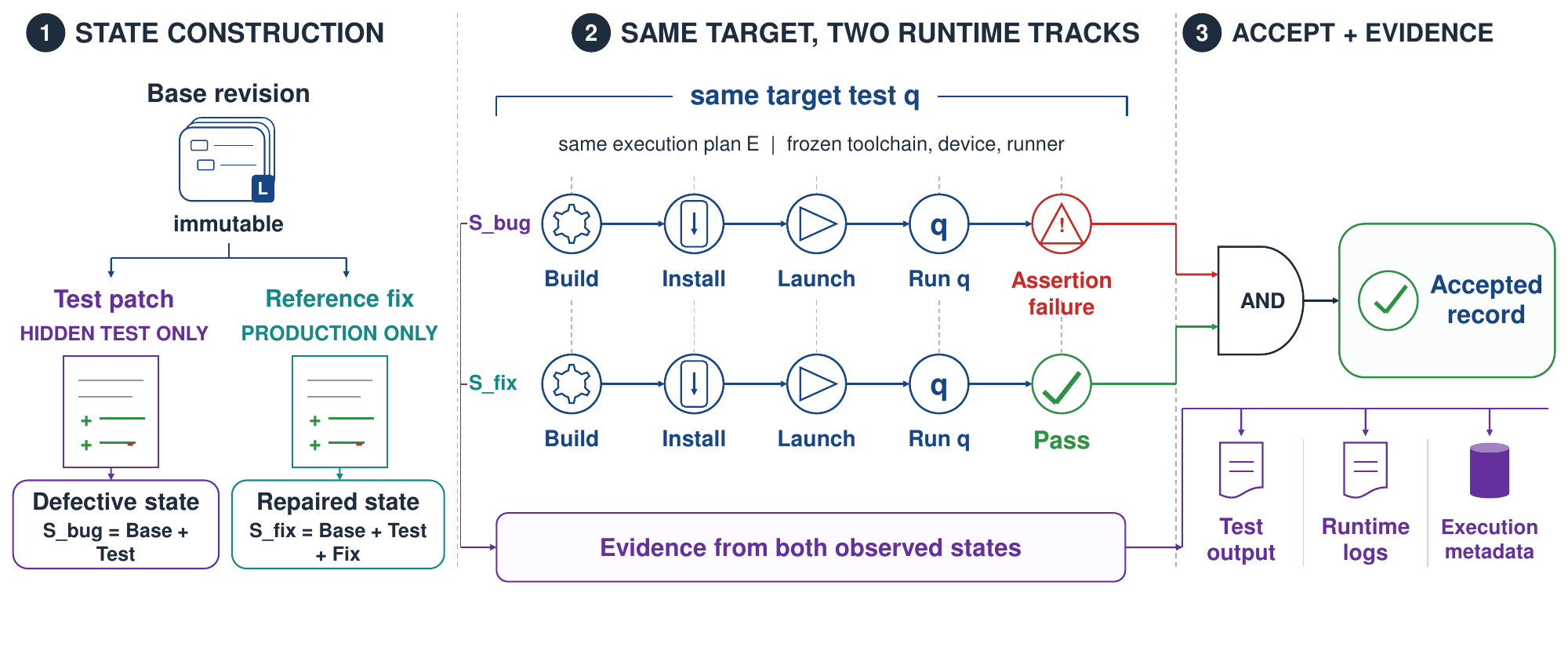}
\caption{Two-state dynamic acceptance. The same target test and frozen
execution plan must reach a behavioral assertion failure on the defective
state and pass on the repaired state. Both executions must preserve complete
runtime evidence.}
\label{fig:two-state}
\end{figure*}

Figure~\ref{fig:two-state} illustrates this two-state acceptance contract.

This definition excludes compilation errors, dependency failures, missing
artifacts, installation errors, crashes before the assertion, application-not-
responding events, offline devices, and timeouts. Those outcomes are retained
as infrastructure evidence and may be retried, but they never satisfy
$A(S_{bug},q)$. The fixed state must run $q$ at least once; a skipped or empty
test selection is not a pass.

\subsection{Localization and Repair Evaluation}

An agent observes $(R,c_b,d)$ and first returns a ranked list of suspected
files $\widehat{F}$, then edits the checked-out repository. Localization is
measured with Hit@$k$, which is one when the first $k$ predictions intersect
$F$, and Recall@$k$, the fraction of $F$ recovered in those predictions.

For repair, the evaluator privately applies $p_t$ and runs $q$ using $E$.
Patch validity records whether a candidate applies cleanly and produces the
required build artifacts. A task counts as fail-to-pass when the candidate
patch builds, installs, launches, and makes $q$ pass. Regression safety records
whether the applicable tests in $Q$ remain passing. We report
$\mathrm{Pass@1}=N_{\mathrm{resolved}}/N_{\mathrm{eligible}}$, where the
eligible set contains completed tasks after excluding unresolved
infrastructure errors. The denominator policy is fixed before evaluation.
Retries may repeat the evaluator on the exact same candidate patch; they do
not grant the agent a new sample.

\section{Benchmark Construction}

\subsection{Repository and Issue Selection}

We target open-source mobile applications with a reproducible build root, a
reachable base revision, and behavior that can be exercised by an installed-
app test. Large sample collections are split only at genuine independent build
roots; support modules remain with the application that needs them. Each
resulting repository is long-lived and carries a provenance manifest linking
it to the upstream URL, path, and source revision.

Candidate defects come from two complementary sources. First, issue- and
pull-request-derived cases preserve real maintenance intent and provide
multi-file or lifecycle-sensitive fixes. Second, controlled behavior mutations
increase coverage of user-visible properties such as text, visibility,
selection, navigation, list contents, persistence, layout, accessibility, and
callback ordering. Controlled mutations are admitted only when a test proves
their behavior on a running application; source-pattern or compilation-only
oracles are insufficient. Every Android record preserves its construction
source so issue-derived and controlled-mutation tasks can be reported
separately.

\subsection{One Issue, One Test}

Each record is mapped to exactly one target method. We first search for an
existing test that directly captures the intended behavior. When none exists,
we add the smallest genuine behavior test using public UI, accessibility
semantics, resources, navigation, state, persistence, or callbacks. Tests that
inspect a newly added test tag or call a fix-only backdoor are rejected. The
one-test rule narrows diagnosis, supports exact runner selection, and prevents
several records from sharing an indistinguishable oracle.

\subsection{Patch Separation and Hygiene}

Starting from a clean base, we create a unique reachable defective commit. The
target test is absent from this commit. We then derive $p_t$ solely from test
restoration or addition and $p_f$ solely from the production correction. Both
patches must apply cleanly to $c_b$; changed paths are inspected, and the
patches are checked for accidental overlap. The record is serialized with a
content hash so that previously generated evidence is reusable only when the
record bytes still match.

\subsection{Verification Procedure}

Algorithm~\ref{alg:verify} summarizes the platform-independent verifier. A
repository lock covers reset, patch application, build, execution, evidence
recording, and cleanup. A device lock covers both states so that no concurrent
task can alter packages, permissions, or runtime state. Each attempt receives
a new artifact directory; failed attempts are never overwritten.

\begin{algorithm}[t]
\caption{Two-state record verification}
\label{alg:verify}
\begin{algorithmic}[1]
\REQUIRE Record $r$, repository lock $\ell_R$, device lock $\ell_D$
\FOR{$(S,o)$ in $[(S_{bug},\mathrm{assert}), (S_{fix},\mathrm{pass})]$}
  \STATE acquire $\ell_R$ and $\ell_D$; reset $R$ to $c_b$
  \STATE apply patches for $S$ and build application/test artifacts
  \IF{build fails}
    \STATE record infrastructure error; restore $R$; release locks; \textbf{continue}
  \ENDIF
  \STATE validate destination; install fresh artifacts; launch application
  \STATE run only target $q$ and preserve runner output and system logs
  \STATE classify assertion, pass, crash, timeout, or infrastructure failure
  \STATE restore repository; release locks
\ENDFOR
\STATE accept iff observed outcomes equal $[\mathrm{assert},\mathrm{pass}]$
\end{algorithmic}
\end{algorithm}

\section{Cross-Platform Execution}

Table~\ref{tab:platforms} shows how the common state machine maps to each
platform. The abstraction deliberately stops at stable semantic operations;
platform adapters retain their native build and test systems.

\begin{table*}[t]
\centering
\small
\begin{tabular}{p{0.09\textwidth}p{0.17\textwidth}p{0.17\textwidth}p{0.20\textwidth}p{0.25\textwidth}}
\toprule
Platform & Build artifacts & Runtime control & Target selection & Behavioral evidence \\
\midrule
Android & App APK + test APK & ADB, emulator/device & instrumentation class\#method & Espresso/UI Automator/runner output \\
iOS & App + XCTest bundle & Xcode simulator/device & \texttt{-only-testing} target & XCTest assertion and simulator log \\
HarmonyOS & HAP + test artifact & HDC, emulator/device & ArkXTest suite/test & ArkXTest assertion and device log \\
\bottomrule
\end{tabular}
\caption{Platform adapters implement the same build--install--launch--run
contract. This paper reports audited quantitative results only for Android.}
\label{tab:platforms}
\end{table*}

\subsection{Android}

The Android adapter checks out the exact base, applies the state patches, and
invokes recorded Gradle tasks with a workspace-local JDK and SDK. It builds the
application and Android-test APKs separately. The verifier binds every ADB
command to one exact serial, validates API level and ABI, waits for boot
completion and a settle interval, wakes and unlocks the device, removes stale
packages, installs fresh APKs, clears application and test data, and invokes
only the recorded instrumentation class and method. Complete instrumentation
output and logcat are preserved. Remote devices are never stopped by a record
verifier; their lifecycle is managed separately.

The final Android construction used API 35 runtime lanes and both ARM64 and
x86-64 devices. Same-repository verification is serialized, while distinct
repositories can run concurrently on separately locked devices. This avoids
two common sources of irreproducibility: overlapping Git resets and ambiguous
ADB device selection.

\subsection{iOS}

The iOS adapter follows the same two-state protocol using an immutable Xcode
project revision, a frozen Xcode/SDK version, and an exact simulator or device
destination. The build step records the scheme, configuration, destination,
derived-data location, application bundle, and XCTest bundle. The execution
step installs and launches a clean application state, then selects a single
XCTest method. Build failures, code-signing failures, unavailable destinations,
simulator boot failures, and pre-assertion crashes are infrastructure errors.
The reconciled dataset will additionally report whether each case uses a
simulator, a physical device, unit-hosted XCTest, or application UI testing.

\subsection{HarmonyOS}

The HarmonyOS adapter records its DevEco/Hvigor toolchain, SDK/API level, HAP
and test artifacts, bundle identifiers, exact HDC target, and ArkXTest target.
The verifier cleans prior installations and application data before each state,
then records test output and device logs. Dependency resolution, signing,
installation, offline target, application crash, and timeout outcomes remain
infrastructure failures. The final manifest will freeze whether each record
runs on an emulator or physical device and will state the exact ArkXTest or
Hypium invocation used by that repository.

\section{Dataset Characteristics}

\subsection{Current Audited Checkpoint}

Table~\ref{tab:dataset} separates audited Android values from synthetic layout
placeholders. Android's canonical dataset has
\AndroidRecords{} records, zero pending records, zero rejected records, and
zero unresolved infrastructure errors in its final audit. Every accepted
Android record is an installed-app instrumentation test.

\begin{table}[t]
\centering
\footnotesize
\begin{tabular}{lrrr}
\toprule
Platform & Accepted & Repositories & Installed-app tests \\
\midrule
Android & \AndroidRecords & \AndroidRepos & \AndroidRecords{} (100\%) \\
iOS & \IOSRecords & \IOSRepos & \IOSRecords{} (100\%) \\
HarmonyOS & \HarmonyRecords & \HarmonyRepos & \HarmonyRecords{} (100\%) \\
\midrule
Total & \TotalRecords & \TotalRepos & \TotalRecords{} (100\%) \\
\bottomrule
\end{tabular}
\caption{Dataset summary. Android is audited; ``sim'' denotes synthetic layout
placeholders.}
\label{tab:dataset}
\end{table}

\subsection{Android Composition}

The \AndroidRecords{} Android records span \AndroidRepos{} independently
buildable repository identities associated with \AndroidOrgs{} upstream
organizations. Their fixes range from one changed line to 49 changed lines.
The median reference-fix size is two changed lines (mean 6.33), while the
median test patch is 28 changed lines (mean 32.35). This distribution reflects
the benchmark's construction strategy: many behavior-level faults are small,
but the tail includes lifecycle, persistence, import/export, permissions,
asynchrony, and multi-file cases.

\begin{table}[t]
\centering
\begin{tabular}{lrrrrr}
\toprule
Patch & Min & Median & Mean & P90 & Max \\
\midrule
Reference fix & 1 & 2 & 6.33 & 18 & 49 \\
Test patch & 5 & 28 & 32.35 & 58 & 99 \\
\bottomrule
\end{tabular}
\caption{Changed-line statistics for the audited Android partition. Counts
include added and deleted lines, excluding diff headers.}
\label{tab:patchsize}
\end{table}

The heavy concentration of small fixes should not be read as evidence that the
tasks are trivial. An agent must still identify the correct production locus,
respect Android lifecycle and resource semantics, and produce a patch that
survives the full build-install-run path. Conversely, the long tail motivates
stratified reporting by patch size and construction source so that aggregate
scores cannot hide weaknesses on more realistic repairs.

\section{Experimental Setup}

\subsection{Research Questions}

Our evaluation is organized around three questions:

\begin{itemize}
    \item \textbf{RQ1: Issue localization.} How accurately do LLMs identify
    defect files across ArkTS, Swift, and Kotlin repositories?
    \item \textbf{RQ2: Automated repair.} How often do generated patches apply,
    achieve fail-to-pass behavior, and preserve regression safety?
    \item \textbf{RQ3: Generalization and failures.} How do results vary by
    platform and complexity, and which failures arise from reasoning, tool use,
    or runtime infrastructure?
\end{itemize}

\subsection{Evaluation Protocol}

The Android experiment evaluates five agent/model configurations on the same
\AndroidEvaluationRecords{} tasks. Each model receives one sample per
task. Evaluation retries, at most two, rerun the exact candidate patch and do
not resample the agent. The evaluator reports a candidate as passed only after
the hidden target test executes successfully. Unresolved infrastructure errors
are excluded from the metric denominator and reported separately.

To quantify whether systems solve the same issues, we encode every eligible
model-task outcome as pass or fail and report Fleiss' $\kappa$ across the five
configurations. For task $i$, category $c$, and $m=5$ systems, agreement is
$P_i=\sum_c n_{ic}(n_{ic}-1)/(m(m-1))$; with global category proportions
$p_c$, we compute $\kappa=(\bar P-\sum_c p_c^2)/(1-\sum_c p_c^2)$. This metric
measures task-level overlap after marginal success rates are accounted for; it
does not replace Pass@1 as the measure of repair quality.

All five Android configurations use the names shown in
Table~\ref{tab:android-results}. Cells marked ``sim'' must be replaced after
matched, dynamically audited iOS and HarmonyOS runs.

\section{Results}

\begin{table}[t]
\centering
\footnotesize
\begin{tabular}{lrrrr}
\toprule
Agent/model & Passed & Failed & Infra. & Pass@1 \\
\midrule
GPT-5.6 Sol & 172 & 28 & 0 & 86.00\% \\
GPT-5.6 Terra & 68 & 132 & 0 & 34.00\% \\
DeepSeek V4 Pro & 84 & 116 & 0 & 42.00\% \\
MiniMax M3 & 44 & 156 & 0 & 22.00\% \\
Qwen3.7-Max & 181 & 19 & 0 & 90.50\% \\
\bottomrule
\end{tabular}
\caption{Verified Android Pass@1 results. Every model is evaluated on all
\AndroidEvaluationRecords{} tasks, and no evaluation has an unresolved
infrastructure error. Across the five binary outcome vectors, Fleiss'
$\kappa=-0.024$.}
\label{tab:android-results}
\end{table}

\begin{table}[t]
\centering
\scriptsize
\setlength{\tabcolsep}{3.5pt}
\begin{tabular}{lccc}
\toprule
Agent/model & Android & iOS & HarmonyOS \\
\midrule
GPT-5.6 Sol & 86.00\% & \IOSGPTSolPass & \HarmonyGPTSolPass \\
GPT-5.6 Terra & 34.00\% & \IOSGPTTerraPass & \HarmonyGPTTerraPass \\
DeepSeek V4 Pro & 42.00\% & \IOSDeepSeekPass & \HarmonyDeepSeekPass \\
MiniMax M3 & 22.00\% & \IOSMiniMaxPass & \HarmonyMiniMaxPass \\
Qwen3.7-Max & 90.50\% & \IOSQwenPass & \HarmonyQwenPass \\
\bottomrule
\end{tabular}
\caption{Cross-platform Pass@1 matrix. Android is verified; ``sim'' denotes
synthetic layout placeholders.}
\label{tab:cross-platform-results}
\end{table}

\subsection{RQ1: Issue Localization}

The Android experiment records final repair outcomes but does not contain
a complete, separately audited localization-output matrix. We therefore defer
Hit@$k$ and Recall@$k$ claims until all model predictions have been reconciled
with the gold defect-file sets. The final experiment will report both micro
averages over tasks and macro averages over repositories for each platform.

\subsection{RQ2: Automated Repair}

Table~\ref{tab:android-results} shows that repository-level agents can solve many
behavior-level Android repairs, but performance differs sharply. The strongest
configuration resolves 181 of 200 tasks (90.50\%), while the weakest resolves
44 (22.00\%). Because all systems use the same task set, denominator
policy, and dynamic oracle, the 68.50-point spread reflects end-to-end system
differences rather than evaluation-policy differences. The aggregate counts do
not by themselves identify whether localization, edit selection, build-feedback
use, or another component causes the gap. Across all five configurations,
549 of 1,000 model-task evaluations pass, giving a 54.90\% micro-average. The
final failure analysis must test explanations for the between-system gap
directly.

Task-level agreement is much weaker than the aggregate success rates suggest.
The observed Fleiss agreement is $\bar P=0.493$, compared with chance agreement
$P_e=0.505$ from the pooled pass/fail marginals, yielding $\kappa=-0.024$.
Thus, the five systems show essentially no agreement beyond chance about which
issues they solve. Because their marginal Pass@1 values differ substantially,
this near-zero $\kappa$ should be interpreted as low overlap among solution
sets, not as a standalone ranking of repair quality.

The high absolute score of the strongest systems is consistent with the
Android partition's many small controlled defects. It does not establish that
mobile repair is solved. First, 19 tasks remain unsolved even for the leading
system. Second, iOS and HarmonyOS introduce different build, signing,
lifecycle, and UI-testing constraints. The final study must therefore report
platform-specific results before making a cross-platform claim.

\subsection{RQ3: Cross-Platform Generalization}

The present evidence supports only a preliminary Android conclusion: small
production diffs can still require repository-level and runtime reasoning. A
proper complexity analysis will stratify records by changed production lines,
changed files, test type, UI interaction depth, asynchronous behavior, and
construction source. We will report macro averages across repositories in
addition to micro averages across records because one large sample collection
contributes many Android cases.

For the cross-platform comparison, raw Pass@1 alone is insufficient. Platform
subsets may differ in defect mix, repository size, and device requirements. We
will therefore provide within-platform scores and a matched analysis over
behavior categories shared by Android, iOS, and HarmonyOS, such as visibility,
navigation, persistence, validation, and callback handling.

\subsection{Failure Modes}

Infrastructure-aware classification is already useful in the Android
experiment: all 1,000 model-task evaluations have a resolved outcome, with
zero unresolved infrastructure errors in the reported matrix. Keeping this
category separate ensures that future runtime failures cannot silently become
incorrect patches and blur the meaning of Pass@1.
The final error analysis will inspect unsuccessful candidate patches and assign
them to mutually exclusive primary causes: no production edit, wrong
localization, incomplete semantic fix, build regression, test-target mismatch,
application crash, timeout, or external infrastructure. We will also report
how often agents edit tests, generated files, or build configuration in ways
that violate the benchmark policy.

\section{Discussion}

\subsection{Why Installed-App Tests Matter}

Host-side unit tests are inexpensive and remain useful, but they cannot observe
many mobile defects. Resource qualifiers, lifecycle restoration, permissions,
intent dispatch, navigation back stacks, accessibility properties, on-device
databases, and UI callbacks depend on the platform runtime. Requiring installed-
app execution makes these behaviors part of the benchmark rather than hidden
environmental assumptions. It also raises the standard for acceptance: a test
that never reaches its assertion cannot validate the defect.

\subsection{Benchmark Integrity}

The evaluator-only representation contains the hidden test patch and reference
fix and must not be released as agent input. Public task material should omit
both and expose only the issue and defective repository revision. Release also
requires per-repository license review, provenance validation, and a leakage
audit. The Android count target is complete, but this manuscript does not claim
that all public-release gates have been satisfied.

Synthetic and issue-derived tasks serve different purposes. Controlled
mutations offer precise oracles and broad behavior coverage, while upstream
issues offer natural descriptions and more realistic fix structure. We retain
source labels and recommend reporting them separately. Future benchmark
versions should expand exact upstream fixes and use time-based splits to reduce
the chance that a model has memorized a public patch.

\subsection{Cost and Reproducibility}

Mobile dynamic evaluation is expensive because every state requires build,
installation, runtime setup, and targeted execution. \BenchmarkName{} controls
this cost through canonical repositories, read-mostly dependency caches,
serial execution per repository, parallelism only across independent
repositories and devices, and hash-based resume. Reproducibility depends less
on a single container image than on fully recording the external runtime:
toolchain versions, destination identity, API/OS version, architecture,
timeouts, package IDs, and logs.

\section{Related Work}

\paragraph{Program repair and executable benchmarks.}
Classical repair studies established generate-and-validate evaluation and the
reuse-oriented plastic-surgery hypothesis
\cite{legoues2012systematic,gazzola2019survey,barr2014plastic}. Reproducible
corpora then broadened the languages and defect sources represented by
executable tasks: Defects4J, ManyBugs/IntroClass, QuixBugs, Codeflaws, Bugs.jar,
BEARS, and BugsInPy cover Java, C, Python, and competition programs
\cite{just2014defects4j,legoues2015manybugs,lin2017quixbugs,tan2017codeflaws,
saha2018bugsjar,madeiral2019bears,widyasari2020bugsinpy}. Repair techniques have
likewise progressed from learned patch ranking and templates to neural
translation and pretrained models
\cite{long2016prophet,liu2019tbar,lutellier2020coconut,xia2022zeroshot,
xia2023plmrepair}. SWE-bench moved the unit of evaluation to repository-level
GitHub issues \cite{jimenez2024swebench}; \BenchmarkName{} retains executable
fail-to-pass evaluation while making app installation, launch, device binding,
and behavior-vs.-infrastructure classification part of correctness.

\paragraph{Repository context and software agents.}
RepoBench and CrossCodeEval isolate whether models can use repository and
cross-file context \cite{liu2023repobench,ding2023crosscodeeval}. SWE-agent,
Agentless, AutoCodeRover, OpenHands, and RepairAgent instead evaluate systems
that search, edit, and execute within a repository
\cite{yang2024sweagent,xia2025agentless,zhang2024autocoderover,
wang2024openhands,bouzenia2025repairagent}. \BenchmarkName{} is agent-agnostic:
it supplies an evaluation contract that exposes mobile-specific failure
boundaries without granting candidate systems access to reference fixes.

\paragraph{Mobile testing and device control.}
Prior Android testing work compared automated input generators and introduced
search-based, model-based, UI-guided, learned, and crash-oriented tools
\cite{choudhary2015android,mao2016sapienz,su2017stoat,li2017droidbot,
li2019humanoid,moran2016crashscope}. AndroidInTheWild provides demonstrations
for device control, while AndroidWorld evaluates task completion in dynamic
apps \cite{rawles2023androidwild,rawles2024androidworld}. These works assess
interaction policies; \BenchmarkName{} instead assesses source repair whose
effect must be observed in an installed application. ArkEval evaluates
repository-level automated repair for ArkTS \cite{xie2026arkeval}, whereas
\BenchmarkName{} places ArkTS repair in a unified mobile protocol spanning
HarmonyOS, iOS, and Android. SolEval similarly values repository context and
executable domain metrics, but targets Solidity smart-contract generation
\cite{peng2025soleval}. Android instrumentation
\cite{androidtesting}, XCTest \cite{applexctest}, and ArkXTest
\cite{openharmonyarkxtest} supply platform runners; our contribution is a
uniform two-state acceptance and evidence model across them.

\section{Threats to Validity}

\paragraph{Construct validity.}
Passing one target test does not prove that a patch is globally correct or
regression-free. We mitigate oracle weakness through behavior-level tests,
patch separation, and manual path review, but future releases should add
broader regression suites where runtime cost permits. Patch size is also an
imperfect proxy for semantic difficulty.

\paragraph{Internal validity.}
Device state, cached dependencies, asynchronous callbacks, and flaky UI tests
can affect outcomes. Exact serial binding, fresh installation and data clearing,
settle intervals, two-state locking, bounded retries, and preserved logs reduce
but do not eliminate this risk. A retry repeats the same agent patch to avoid
conflating evaluator recovery with additional model sampling.

\paragraph{External validity.}
The current audited statistics are dominated by Android and by sample-project
build roots. Results may not generalize to large commercial applications,
physical-device-only features, or all OS versions. The iOS and HarmonyOS
partitions and additional upstream fixes are necessary to test whether findings
transfer across toolchains and application frameworks.

\paragraph{Data leakage.}
Some tasks derive from public issues or pull requests and may occur in model
training data. Controlled mutations reduce direct overlap but may simplify the
task distribution. We therefore preserve provenance, recommend separate scores
by construction source, and plan temporal and exact-patch overlap analyses.

\section{Ethics and Artifact Policy}

The benchmark uses open-source repositories and should preserve their licenses
and attribution requirements. Public artifacts must exclude secrets, signing
credentials, personal device data, and evaluator-only reference fixes. Device
logs can contain package names, file paths, or incidental text; they require
screening before release. We intend \BenchmarkName{} for research on software
maintenance and evaluation, not for bypassing platform security or publishing
unreviewed patches to upstream projects.

All reported empirical numbers must be traceable to immutable JSONL records,
audit summaries, and complete state logs. Candidate, pending, rejected, and
infrastructure-error counts should be reported separately. In particular, the
``sim''-marked iOS and HarmonyOS values in this draft are intentionally
synthetic and must be replaced, not cited, before submission.

\section{Conclusion}

We presented \BenchmarkName, a benchmark and evaluation contract for
repository-level mobile repair across HarmonyOS/ArkTS, iOS/Swift, and
Android/Kotlin. Its core requirement is strict: the same installed-app target
must reach a behavior assertion failure on the defective state and pass on the
fixed state, while infrastructure failures remain separate. The audited Android
partition contains \AndroidRecords{} accepted instrumentation tasks across
\AndroidRepos{} buildable repositories. Across these tasks, five agents range
from 22.00\% to 90.50\% Pass@1, demonstrating that the choice
of agent materially affects end-to-end repair success even on one platform.
These findings are not yet evidence of cross-platform generalization. That
claim requires audited iOS and HarmonyOS partitions, matched agent protocols,
and the planned localization and stratified failure analyses.

\bibliography{references}

\end{document}